\documentstyle[12pt,epsfig]{ifae}
\newcommand{\bra}[1]{\left\langle #1 \right|}
\newcommand{\ket}[1]{\left| #1 \right\rangle}
\begin{document}


\begin{header}
  \title{ \begin{center}
  Interpretation of \\
D$_{sJ}^*$(2317) and D$_{sJ}$(2460)
          \end{center}}

  \begin{Authlist}
    Rossella~Ferrandes

  Dipartimento di Fisica, Universit\`a degli Studi di Bari, Bari, Italy
  \end{Authlist}

  \begin{abstract}
    The experimental status of the recently observed resonances which could be interpreted as c$\overline q$ mesons with s$_l^P=\frac{1}{2}^+$
  is reviewed. In the framework of HQET and chiral perturbation theory, strong and radiative widths of these states are computed in the
  hypothesis that they are c$\overline q$ states, obtaining results consistent with the experimental measurements. Masses and widths for the analogous
  states containing a beauty quark are also predicted.
  \end{abstract}

\end{header}

\section{Introduction}
Let us consider a meson containing a single heavy quark Q. In the
infinite heavy quark mass limit the heavy quark spin $S_Q$ and the
total angular momentum of light degrees of freedom $s_l=L+S_q$
decouple. Therefore Q$\overline{q}$ mesons can be classified in
doublets labelled by $s_l$ and parity eigenvalues, the members of
which are degenerate in mass. In the lowest lying states the
orbital angular momentum of light degrees of freedom is L=0 and it
results $s_l=\frac{1}{2}$. The corresponding doublet is composed
by two states with $J^P_{s_l}=(0^-,1^-)_{\frac{1}{2}}$. For L=1 we
have the $s_l^P=\frac{1}{2}^+$ and $s_l^P=\frac{3}{2}^+$ doublets,
with $J^P_{s_l}=(0^+,1^+)_{\frac{1}{2}}$ and
$J^P_{s_l}=(1^+,2^+)_{\frac{3}{2}}$, respectively.

Let us focus on the case of charmed mesons and denote the members
of the positive parity doublets as:
$(0^+,1^+)_{\frac{1}{2}}=(D_{(s)0}^*, D_{(s)1}')$ and
$(1^+,2^+)_{\frac{3}{2}}=(D_{(s)1}, D_{(s)2}^*)$. Conservation of
parity, total angular momentum and heavy quark spin require that
the strong decays
D($s_l^P=\frac{3}{2}^+$)$\rightarrow$D$^{(*)}\pi$ proceed via a
d-wave transition while the decays
D($s_l^P=\frac{1}{2}^+$)$\rightarrow$D$^{(*)}\pi$ proceed via an
s-wave transition. As a consequence, the states with
$s_l^P=\frac{1}{2}^+$ are expected to be broad, whereas those with
$s_l^P=\frac{3}{2}^+$ are presumably narrow.

Last year some resonances were discovered which could be
interpreted as $s_l^P=\frac{1}{2}^+$ states of c$\overline{u}$ and
c$\overline{s}$ systems. These states constitute the subject of
this talk. The non strange resonances are detected with broad
width, as expected, whereas those with strangeness are observed to
be very narrow. The discrepancy has given rise to various
interpretations. After a description of the recent experimental
observations we shall discuss several interpretations of the
detected resonances through the analysis of their decay modes.

\section{Experimental observations}

\subsection{Broad $c\overline{u}$ states}
\begin{figure}
 \begin{center}
  \includegraphics[width=0.32\textwidth] {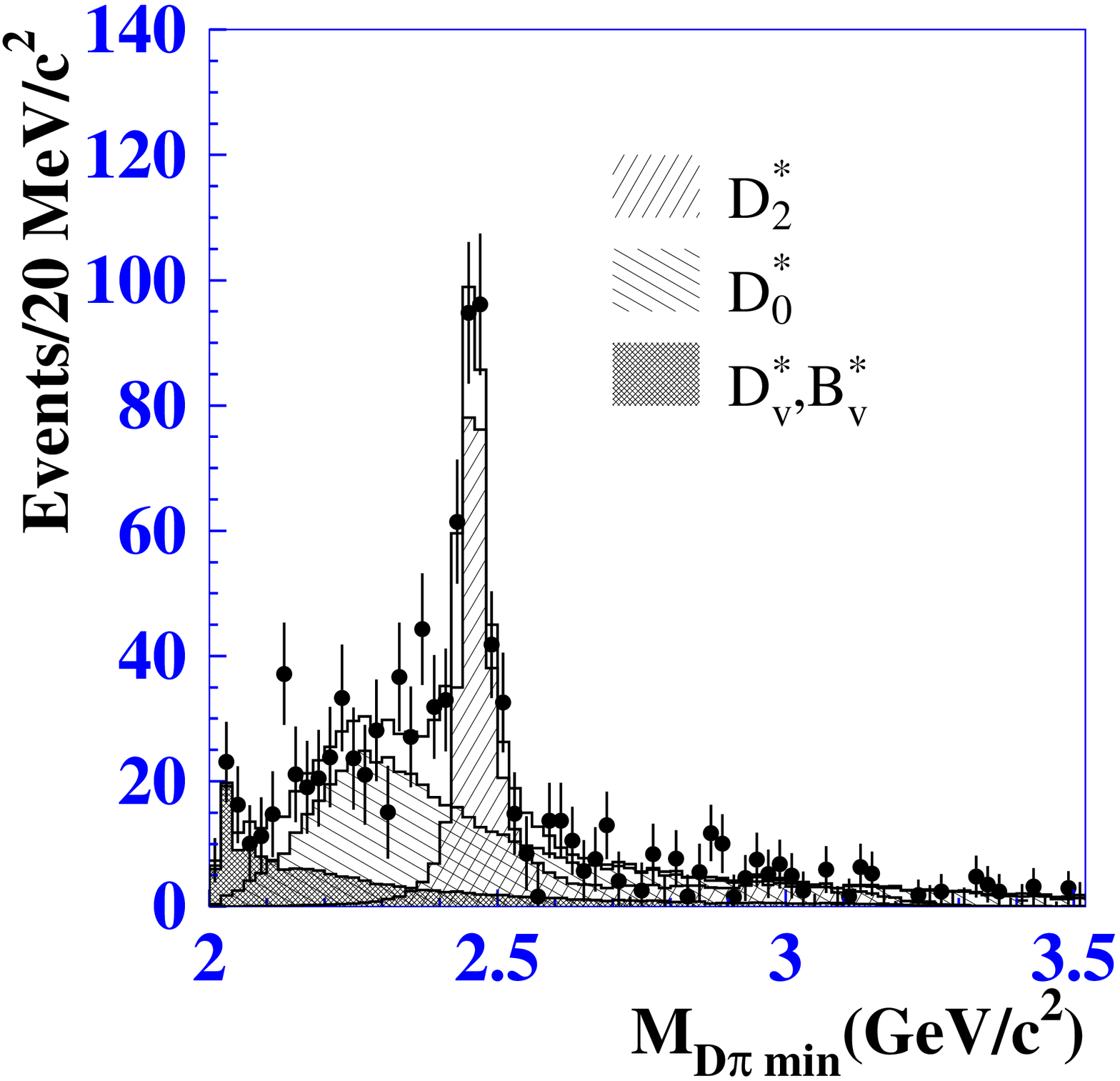}\hspace{1cm}
  \includegraphics[width=0.32\textwidth] {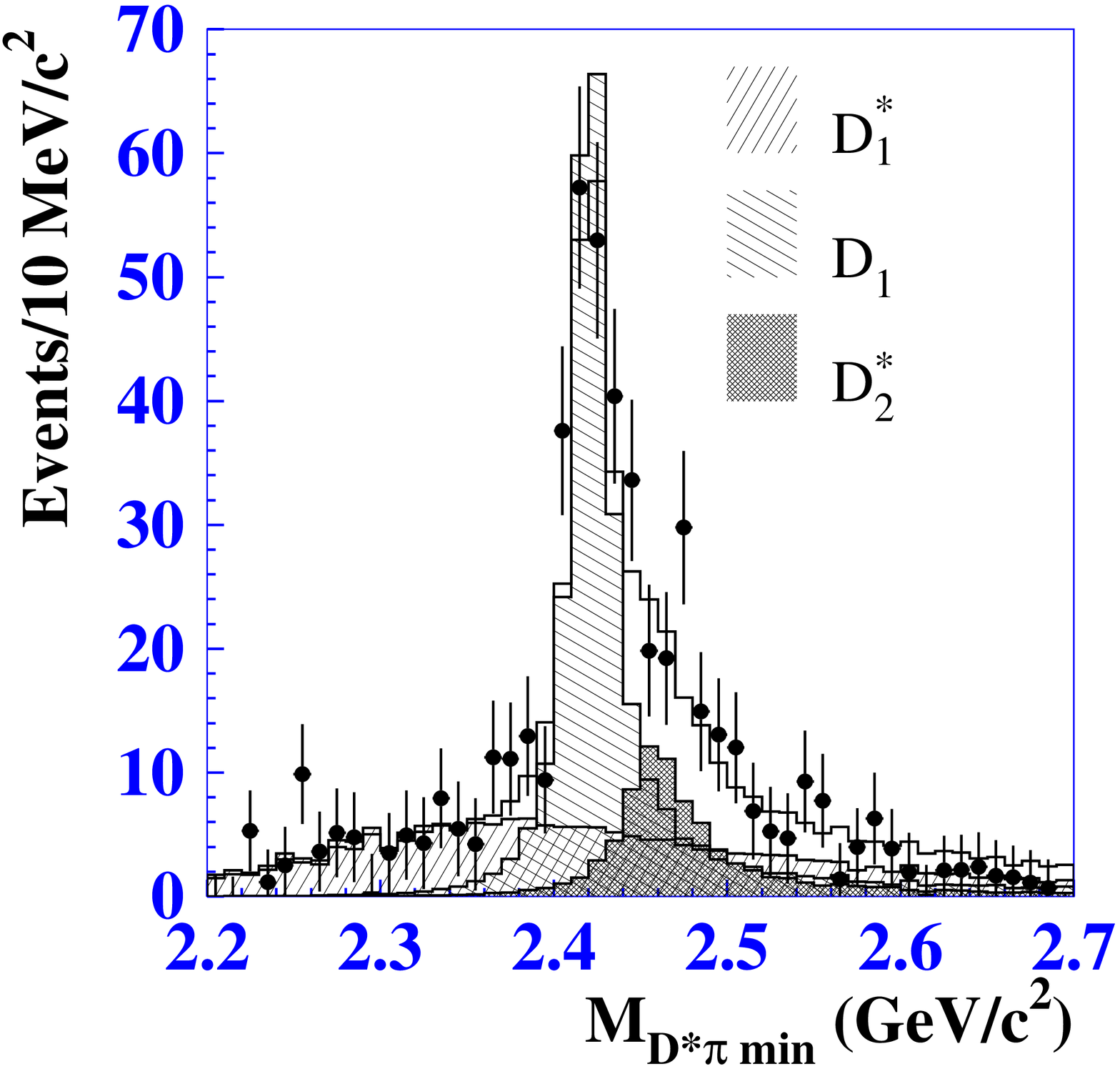}
\vspace*{0mm}
 \caption{The background-subtracted minimal D$\pi$ (left) and D$^*\pi$ (right) mass distributions observed by Belle~\cite{Bellelarghi}. Hatched histograms show different contributions, the open histogram shows the coherent sum of all contributions.}
  \label{Bellelarghi}
 \end{center}
\end{figure}

Last year Belle observed two broad resonances containing a charm
quark which could be identified as $s_l^P=\frac{1}{2}^+$
$c\overline{u}$ states~\cite{Bellelarghi}. These resonances are
observed in B$\rightarrow D^{**}\pi$ decays (where D$^{**}$
indicates $L$=1 states) which have been studied using the
D$^+\pi^-\pi^-$ and D$^{*+}\pi^-\pi^-$ final states. The Dalitz
plot analysis carried out for the D$\pi\pi$ final state includes
the amplitude of the known D$_2^{*0}\pi^-$ mode, possible
contributions of the processes with virtual D$^{*0}\pi^-$ and
B$^{*0}\pi^-$ production and an intermediate D$\pi$ broad
resonance with free mass and width and assigned J$^P=0^+$ quantum
numbers. A fit of the projection of the Dalitz plot to the $D\pi$
axis, where the pion is  the one having the smallest momentum,
favours the presence  of the scalar contribution. A similar
analysis is performed for the D$^{*+}\pi^-\pi^-$ final state,
obtaining the evidence for a broad resonance with J$^P=1^+$
quantum numbers. The D$\pi$ and D$^*\pi$ mass distributions are
shown in Fig.~\ref{Bellelarghi}, whereas the mass and width of the
broad states obtained by the fit are collected in
Table~\ref{tab:statilarghi}.

An analogous study has been carried out by FOCUS Collaboration
\cite{focus}, which considered both the $D^0 \pi^+$ and $D^+\pi^-$
charge configurations.
 Also in this case, a broad scalar contribution
is requested in order to fit the $D \pi$  mass distribution.  The
values of mass and width quoted by FOCUS are collected  in
Table~\ref{tab:statilarghi}. Although the values obtained for the
mass of the scalar state are rather different, presumably as a
consequence of the difficulty in fitting a scalar component, we
include in Table~\ref{tab:statilarghi} the average of Belle and
FOCUS data.

\begin{table}[htbp]
    \caption{Masses and widths of broad resonances observed in D$\pi$ and
D$^*\pi$ systems.}
    \label{tab:statilarghi}
    \begin{center}
    \begin{tabular}{ccccc}
      \hline
      \ & \ &  Belle~\cite{Bellelarghi} &   FOCUS~\cite{focus}  & Average\\
      \hline
      \ D$_0^{*0}$ & \begin{tabular}{c} M (MeV) \\ $\Gamma$ (MeV)
\end{tabular}   &
      \begin{tabular}{c} $2308\pm17\pm15\pm28$ \\ $276\pm21\pm18\pm60$
\end{tabular}   &
      \begin{tabular}{c}  $2407\pm21\pm35$ \\ $240\pm55\pm59$ \end{tabular} &
      \begin{tabular}{c}  $2351\pm27$ \\ $262\pm51$ \end{tabular}
\\ \hline
         \ D$_0^{*+}$ & \begin{tabular}{c} M (MeV) \\ $\Gamma$ (MeV)
\end{tabular}   &
      \begin{tabular}{c} $$ \\ $$
\end{tabular}   &
      \begin{tabular}{c}  $2403\pm14\pm35$ \\ $283\pm24\pm34$ \end{tabular} &
      \begin{tabular}{c}  $$ \\ $$ \end{tabular}
\\

      \hline \hline
      \ D$_1'^{0}$ & \begin{tabular}{c} M (MeV) \\ $\Gamma$ (MeV)
\end{tabular}   &
      \begin{tabular}{c} $2427\pm26\pm20\pm15$\\
                                       $384_{-75}^{+107}\pm24\pm70$\end{tabular} &
                                                         &
\\
      \hline
    \end{tabular}
  \end{center}
\end{table}

\subsection{D$_{sJ}^*$(2317) and D$_{sJ}$(2460)}

Two narrow states have been observed with charm and strangeness,
which could be interpreted as s$_l^P=\frac{1}{2}^+$ states.

\begin{figure}
 \begin{center}
  \includegraphics[width=0.31\textwidth] {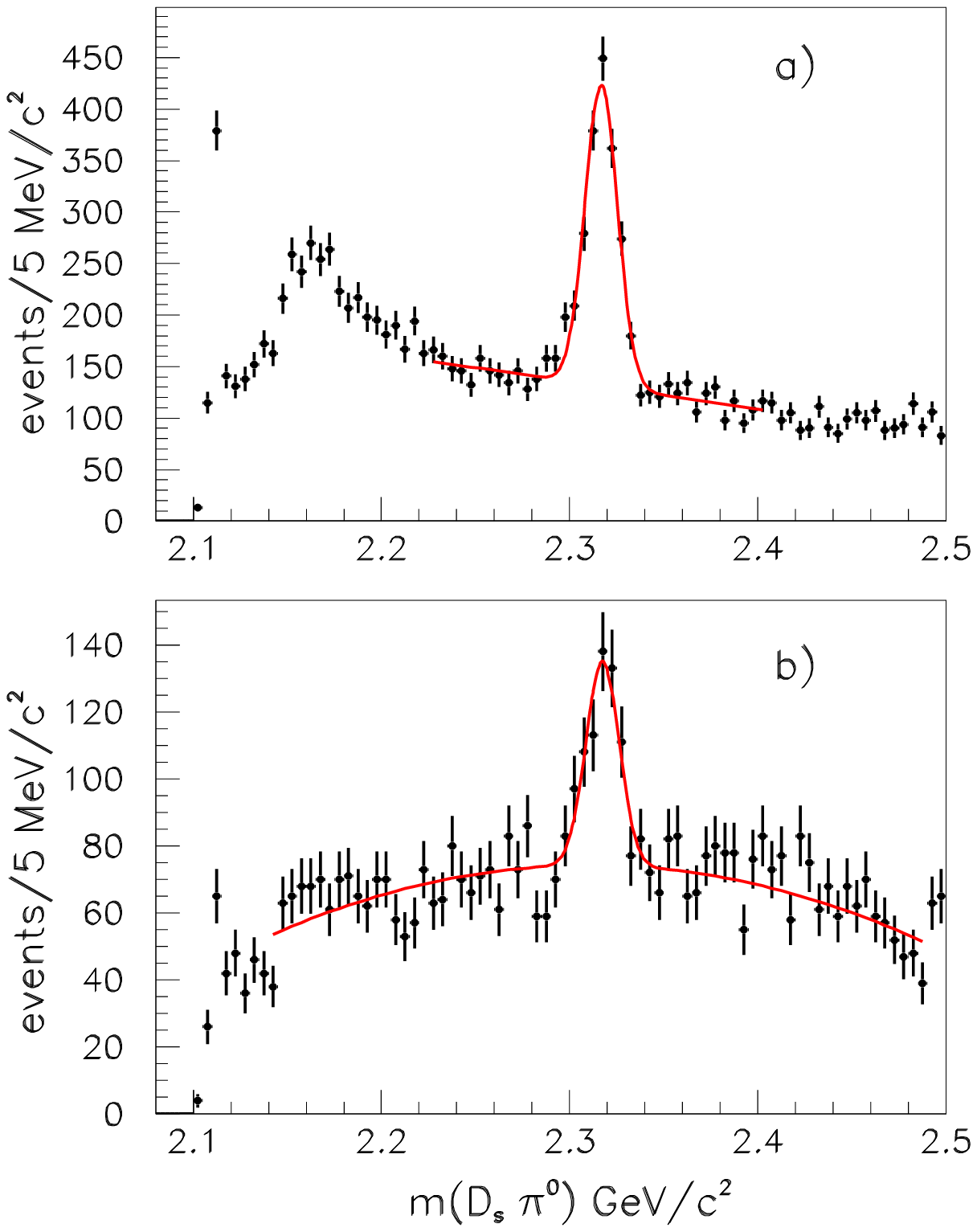}\hspace{1cm}
  \includegraphics[width=0.342\textwidth] {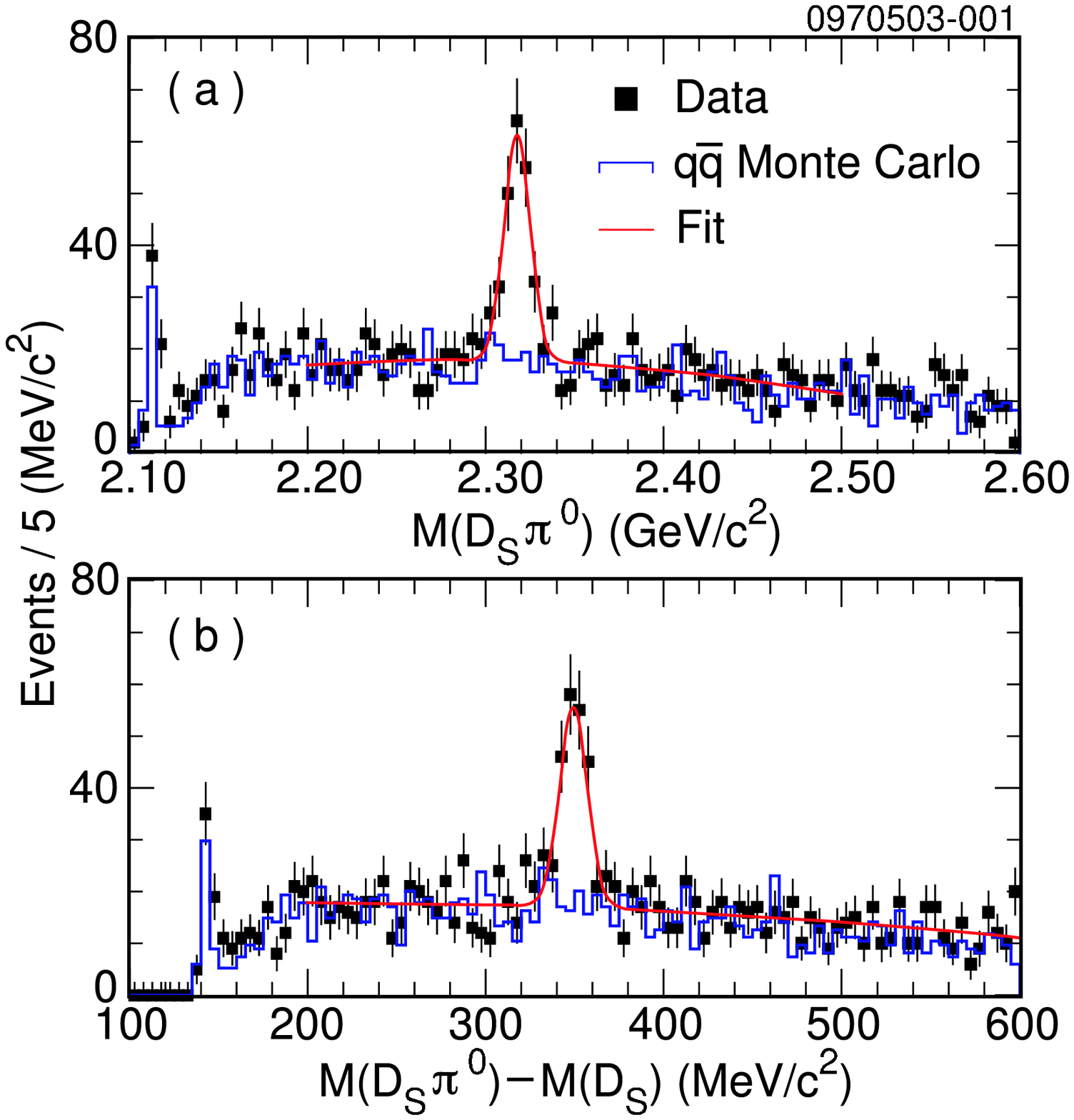}
\vspace*{0mm}
 \caption{Left: The D$_{s}^+\pi^0$ mass distribution for (a) the decay D$_s^+\rightarrow K^+K^-\pi^+$ and (b) the decay D$_s^+\rightarrow K^+K^-\pi^+\pi^0$
 observed by BaBar~\cite{BaBar2317}. Right: Distribution of (a) the masses M(D$_s\pi^0$) and (b) the mass differences $\Delta M(D_s\pi^0)=M(D_s\pi^0)-M(D_s)$ for the D$_s\pi^0$ candidates in the decay D$_s^+\rightarrow K^+K^-\pi^+$, measured by CLEO~\cite{CLEO}.}
  \label{Dsj2317}
 \end{center}
\end{figure}

BaBar Collaboration discovered the narrow resonance
D$_{sJ}^*$(2317) in the D$_s\pi^0$ system~\cite{BaBar2317}, with
mass close to 2.32~ GeV and width consistent with the experimental
resolution ($\sim$10 MeV). This resonance, shown in
Fig.~\ref{Dsj2317}, was confirmed by Belle~\cite{Belle continuo},
CLEO~\cite{CLEO} and recently by FOCUS~\cite{FOCUS2317}, with
similar properties. FOCUS preliminary measurement of the mass is
$M= 2323\pm~2$~MeV, slightly above the values reported in
Table~\ref{tab:Dsj}.

\begin{figure}[hbt]
  \begin{center}
    \epsfig{file=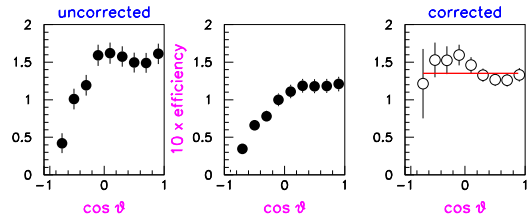,width=0.5\linewidth}
    \caption{BaBar helicity angle analysis of the decay D$_{sJ}^*$(2317)$\rightarrow$D$_s\pi^0$ \cite{Porter}. Left: Uncorrected angular distribution.
    Center: Dependence of efficiency on angle. Right: Efficiency-corrected angular distribution.}
    \label{fig:babar}
  \end{center}
\end{figure}

The occurence of the decay in D$_s\pi^0$ implies that
D$_{sJ}^*$(2317) has natural spin-parity. Furthermore BaBar
helicity analysis for this decay is consistent with spin 0
assignment to D$_{sJ}^*$(2317) (Fig.~\ref{fig:babar}), though such
analysis does not rule out different possibilities since the same
distribution could also result from an isotropic production
polarization. Further support to the $J^P=0^+$ hypothesis is the
absence of a peak in the D$_s\gamma$ system.

\begin{figure}
 \begin{center}
  \includegraphics[width=0.31\textwidth] {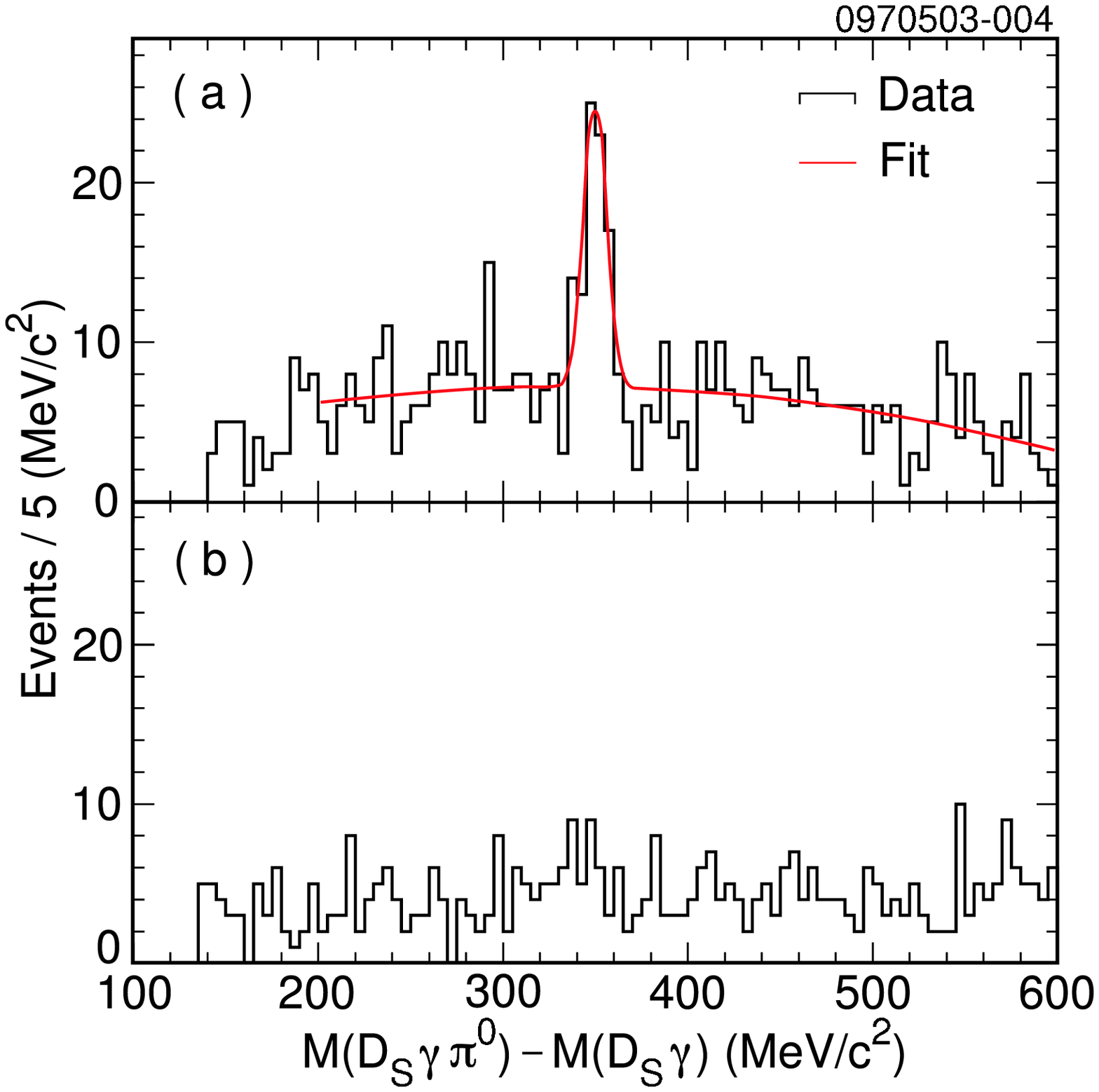}\hspace{1.5cm}
  \includegraphics[width=0.31\textwidth] {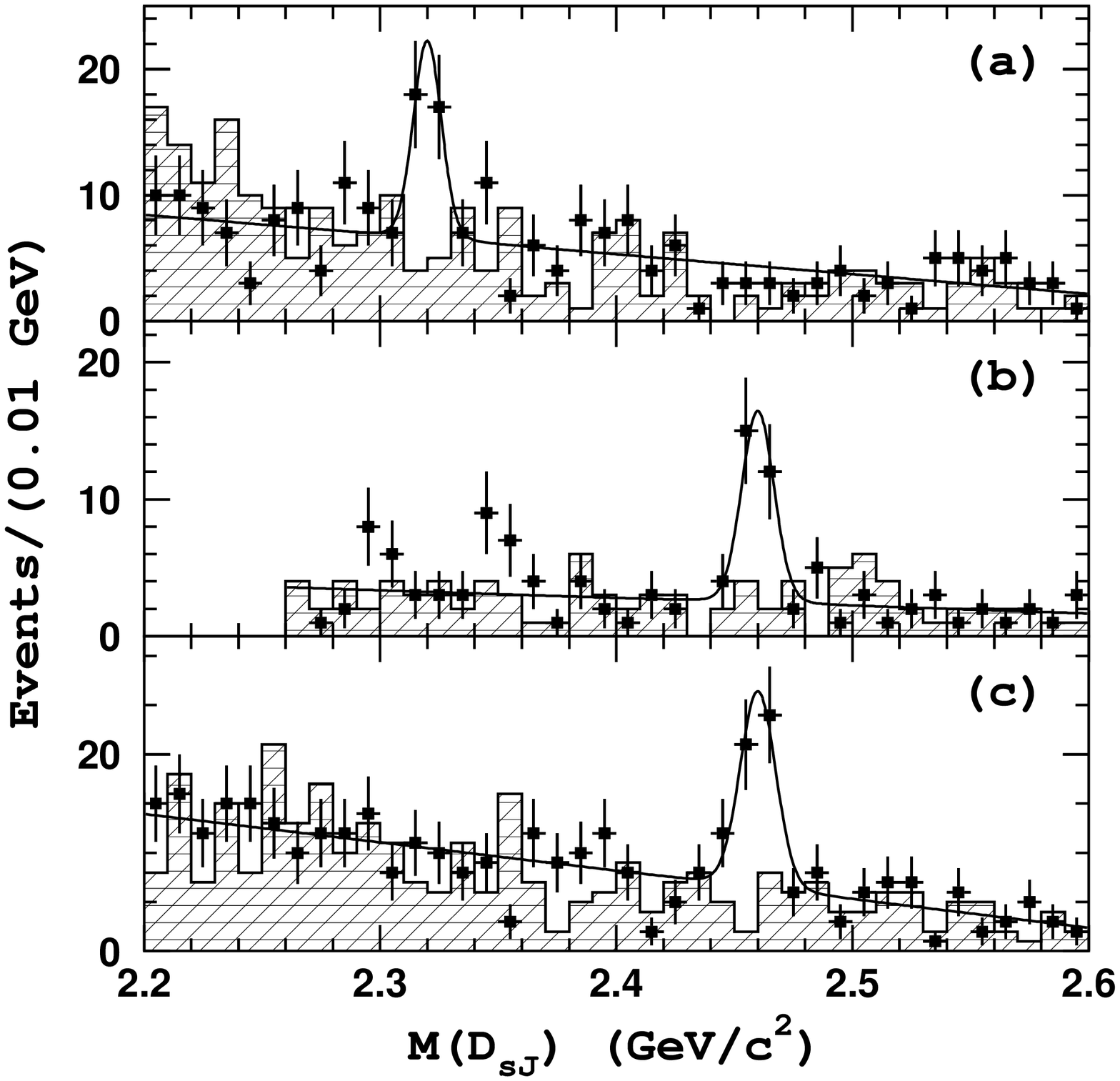}
\vspace*{-2mm}
 \caption{Left: The mass difference spectrum $\Delta M(D_s^*\gamma\pi^0)-M(D_s\gamma)$ measured by CLEO~\cite{CLEO} (a) for combinations where
 the D$_s\gamma$ system is consistent with D$_s^*$ decay and (b) for D$_s\gamma$ combination selected from the D$_s^*$ side band regions. Right: M(D$_{sJ}$) distribution for the
 B$\rightarrow\overline{D}D_{sJ}$ candidates measured by Belle\cite{Belle dal B}: (a) D$_{sJ}(2317)\rightarrow D_s\pi^0$, (b) D$_{sJ}(2460)\rightarrow D_s^*\pi^0$ and (c) D$_{sJ}(2460)\rightarrow D_s\gamma$ }
  \label{Dsj2460}
 \end{center}
\end{figure}

CLEO discovered a narrow resonance in the D$_s^*\pi^0$
system~\cite{CLEO}, named D$_{sJ}(2460)$, soon confirmed by
BaBar~\cite{BaBar2460} and Belle~\cite{Belle continuo}. The
experimental values for mass and width are shown in
Table~\ref{tab:Dsj}. Belle also reports the first observation of
the radiative decay $D_{sJ}(2460)\rightarrow D_s\gamma$
(Fig.~\ref{Dsj2460}), soon confirmed by BaBar. Belle and BaBar
have observed the production of the $D_{sJ}$ states also from B
decays~\cite{Belle dal B}-\cite{BaBardalB}, detecting clean
signals for $B\rightarrow \overline{D} D_{sJ}^*(2317)[D_s\pi^0]$,
$B\rightarrow \overline{D} D_{sJ}(2460)[D_s^*\pi^0]$ and
$B\rightarrow \overline{D} D_{sJ}(2460)[D_s\gamma]$ channels.
Furthermore BaBar reports the observation of the analogous decay
modes involving combinations of the $D_{sJ}$ with a neutral or
charged $D^*$. The measured branching ratios are shown in
Table~\ref{tab:BR}.

The ratio $\frac{{\mathcal B}(D_{sJ}(2460)\rightarrow
D_s\gamma)}{{\mathcal B}(D_{sJ}(2460)\rightarrow D_s^*\pi^0)}$
obtained by Belle from continuum analysis is 0.55$\pm 0.13
\pm0.08$, whereas the corresponding value from B decays is
0.38$\pm 0.11 \pm0.04$. BaBar preliminary analysis does not report
a value for this quantity, yet. However using its measurements of
single branching ratios and assuming that statistical and
systematic errors are independent for each channel and among
different channels, one obtains $\frac{{\mathcal
B}(D_{sJ}(2460)\rightarrow D_s\gamma)}{{\mathcal
B}(D_{sJ}(2460)\rightarrow D_s^*\pi^0)}=0.44\pm0.17$, consistently
with Belle measurements.

\begin{table}[htbp]
    \caption{Masses and widths of the narrow resonances D$_{sJ}^*(2317)$ and
D$_{sJ}(2460)$ measured by BaBar, Belle and CLEO Collaborations.
The average value for the mass is also reported.}
 \label{tab:Dsj}
     \begin{center}
    \begin{tabular}{ccc}
    \\ \hline
      $D^*_{sJ}(2317)$ &   $D_{sJ}(2460)$ &  ref.\\  \hline
      \begin{tabular}{cc}
      M (GeV) & $\Gamma$ (GeV) \\  \hline
      $2317.3\pm0.4\pm0.8$ & $<10$ \\
      $2317.2\pm0.5\pm0.9$ & $<4.6$ \\
      $2318.5\pm1.2\pm1.1$ & $<$7 \\
      $2317.4\pm0.6$ &$$
      \end{tabular}
      &
       \begin{tabular}{cc}
       M (GeV) & $\Gamma$ (GeV) \\ \hline
      $2458.0\pm1.0\pm1.0$ & $<10$ \\
      $2456.5\pm1.3\pm1.3$ & $<5.5$ \\
      $2463.6\pm1.7\pm1.2$ & $<$7 \\
      $2458.8\pm1.0$ &$$
             \end{tabular}
              &
       \begin{tabular}{c}
       BaBar~\cite{BaBar2317}-\cite{BaBar2460} \\
       Belle~\cite{Belle dal B}             \\
       CLEO~\cite{CLEO} \\
        \end{tabular}       \\
       \hline
      \end{tabular}
  \end{center}
\end{table}


The D$_{sJ}(2460)\rightarrow D_s^*\pi^0$ decay implies that
D$_{sJ}(2460)$ has unnatural spin-parity. The observation of its
radiative decay in D$_s\gamma$ rules out the possibility that it
is J=0, and helicity distribution measured by Belle and BaBar in B
decays is consistent with J=1. These arguments supports J$^P=1^+$
hypothesis.

\begin{figure}[hbt]
 \begin{center}
  \includegraphics[width=0.27\textwidth] {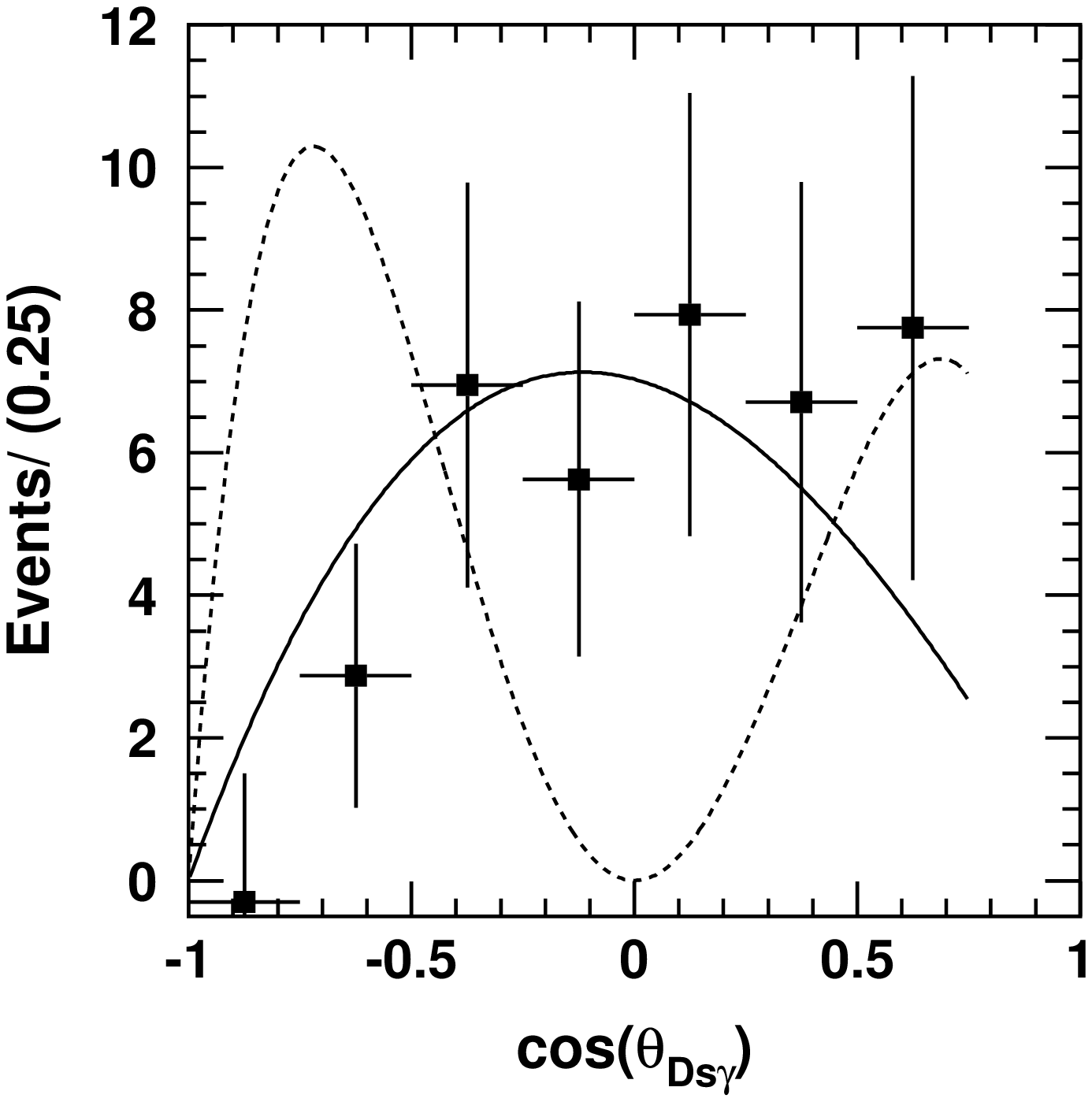}\hspace{1.2cm}
  \includegraphics[width=0.26\textwidth] {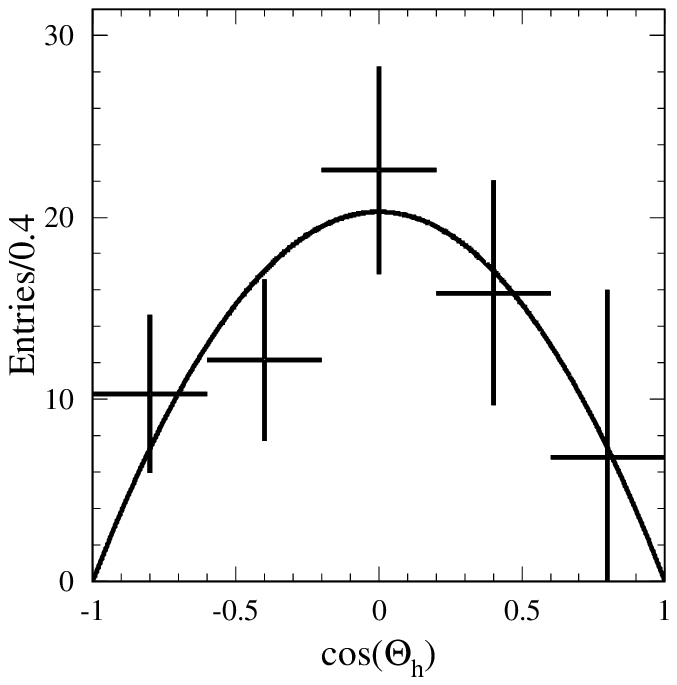} \hspace{-0.4cm}
  \includegraphics[width=0.26\textwidth] {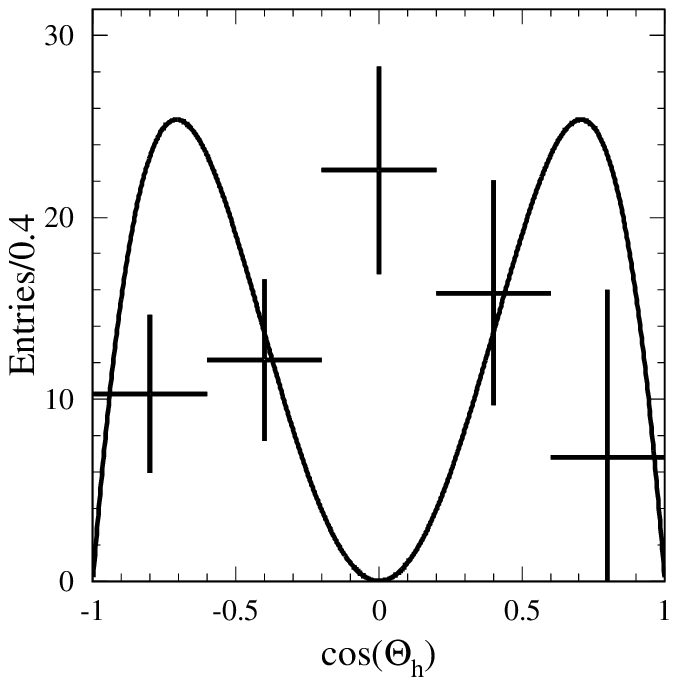}
 \caption{Helicity distribution of  D$_{sJ}(2460)\rightarrow D_s\gamma$
measured  by Belle~\cite{Belle dal B} (left) and
BaBar~\cite{BaBardalB} (right).  The distributions are consistent
with the assignment $J=1$ (continuous line in the left panel,
first plot in the right panel), and not with $J=2$
 (dashed line in the left panel, second plot in the right panel).}
\label{fig:belleangolare}
 \end{center}
\end{figure}

\begin{table}[h]
\caption{Branching fractions $(10^{-3})$ measured by BaBar and
Belle. Upper limits (at 90\% C.L.) are shown in parentheses.}
\label{tab:BR}
\begin{center}
\begin{tabular}{c c c c}
 \hline
Decay Mode & BaBar~\cite{BaBardalB} & Belle~\cite{Belle dal B}& Average \\
    \hline
  $B^0\rightarrow D_{s0}^{*}D^{-}$ $(D_{s0}^{*}\rightarrow
D_{s}^{+} \pi^0)$  &
 $2.09 \pm 0.40\pm 0.34 ^{+0.70}_{-0.42}$  & $0.86\pm0.26{}^{+0.33}_{-0.26}$
 &$1.09\pm0.38$
\\
  $B^0\rightarrow D_{s0}^{*}D^{*-}$ $(D_{s0}^{*}\rightarrow
D_{s}^{+} \pi^0)$
& $1.12 \pm 0.38\pm 0.20^{+0.37}_{-0.22}  $ & \ \hfill --- \hfill\ \\
  $B^+\rightarrow D_{s0}^{*}\overline{D}^{0}$
$(D_{s0}^{*}\rightarrow  D_{s}^{+} \pi^0)$ & $1.28 \pm 0.37\pm
0.22 ^{+0.42}_{-0.26}$ & $0.81\pm0.24{}^{+0.30}_{-0.27}$ & $0.94
\pm 0.32$
 \\
  $B^+\rightarrow D_{s0}^{*}\overline{D}^{*0}$
$(D_{s0}^{*}\rightarrow  D_{s}^{+} \pi^0)$
& $1.91 \pm 0.84\pm 0.50 ^{+0.63}_{-0.38}$ & \ \hfill --- \hfill\ \\
  \hline
  $B^0\rightarrow D_{s0}^{*}D^{-}$ $(D_{s0}^{*}\rightarrow
D_{s}^{*+} \gamma)$  &
 \ \hfill --- \hfill\  & $0.27^{+0.29}_{-0.22}(<0.95)$\\
  $B^+\rightarrow D_{s0}^{*}\overline{D}^{0}$
$(D_{s0}^{*}\rightarrow  D_{s}^{*+} \gamma)$ &  \ \hfill ---
\hfill\ & $0.25^{+0.21}_{-0.16}(<0.76)$ \\ \hline

 $B^0\rightarrow D_{s1}'D^{-}$ $(D_{s1}'\rightarrow
D_{s}^{*+} \pi^0)$ & $1.71 \pm 0.72\pm 0.27 ^{+0.57}_{-0.35}$ &
$2.27\pm0.68{}^{+0.73}_{-0.62}$ & $1.98 \pm0.69$
\\
 $B^0\rightarrow D_{s1}'D^{*-}$ $(D_{s1}'\rightarrow
D_{s}^{*+} \pi^0)$
& $5.89 \pm 1.24 \pm 1.16^{+1.96}_{-1.17} $ & \ \hfill --- \hfill\ \\
  $B^+\rightarrow D_{s1}'\overline{D}^{0}$
$(D_{s1}'\rightarrow  D_{s}^{*+} \pi^0)$ & $2.07 \pm 0.71\pm 0.45
^{+0.69}_{-0.41}$ & $1.19\pm0.36{}^{+0.61}_{-0.49}$ & $1.45\pm
0.59$
\\
  $B^+\rightarrow D_{s1}'\overline{D}^{*0}$
$(D_{s1}'\rightarrow  D_{s}^{*+} \pi^0)$
& $7.30 \pm 1.68 \pm 1.68^{+2.40}_{-1.43}$ & \ \hfill --- \hfill\ \\
  \hline
  $B^0\rightarrow D_{s1}'D^{-}$ $(D_{s1}'\rightarrow
D_{s}^{+} \gamma)$ & $0.92 \pm 0.24\pm 0.11 ^{+0.30}_{-0.19}$ &
$0.82\pm0.25{}^{+0.22}_{-0.19}$ & $0.86 \pm 0.25$
\\
  $B^0\rightarrow D_{s1}'D^{*-}$ $(D_{s1}'\rightarrow
D_{s}^{+} \gamma)$
& $2.60 \pm 0.39   \pm 0.34^{+0.86}_{-0.52}$ & \ \hfill --- \hfill\ \\
  $B^+\rightarrow D_{s1}'\overline{D}^{0}$
$(D_{s1}'\rightarrow  D_{s}^{+} \gamma)$ & $0.80 \pm 0.21\pm 0.12
^{+0.26}_{-0.16}$ & $0.56\pm0.17{}^{+0.16}_{-0.15}$  & $0.63\pm
0.19$
\\
  $B^+\rightarrow D_{s1}'\overline{D}^{*0}$
$(D_{s1}'\rightarrow  D_{s}^{+} \gamma)$
& $2.26 \pm 0.47\pm 0.43 ^{+0.74}_{-0.44}$ & \ \hfill --- \hfill\ \\
  \hline
  $B^0\rightarrow D_{s1}'D^{-}$ $(D_{s1}'\rightarrow
D_{s}^{*+} \gamma)$
& \ \hfill --- \hfill\ & $0.13^{+0.20}_{-0.14}(<0.6)$\\
  $B^+\rightarrow D_{s1}'\overline{D}^{0}$
$(D_{s1}'\rightarrow  D_{s}^{*+} \gamma)$ & \ \hfill --- \hfill\ &
$0.31^{+0.27}_{-0.23}(<0.98)$\\ \hline
\end{tabular}
\end{center}
\end{table}

\section{Interpretations}

Though states decaying via s-wave are expected to be broad, the
strange resonances D$_{sJ}^*(2317)$ and D$_{sJ}(2460)$ have been
detected with narrow widths. Such narrowness can be explained in
the c$\overline{s}$ interpretation with the isospin violation
occurring in the observed decays, which are the only two body
strong decays kinematically allowed according to experimental
masses. However, some potential models predict for these states
masses above the thresholds of the isospin conserving decays in DK
and D$^*$K, respectively for the 0$^+$ and 1$^+$ states, so that
these states were expected as broad resonances in these
systems~\cite{modelli a potenziale}. Since potential models
resulted rather accurate in reproducing the spectrum of other
charmed mesons, this discrepancy has caused different
interpretations to be worked out. Barnes, Close and Lipkin propose
a DK molecular state~\cite{molecola DK}; similarly, Szczepaniak
suggests a D$\pi$ atom~\cite{atomo Dpi}. H. Y. Cheng and W. S. Hou
propose a 4-quark state~\cite{Cheng} and Browder, Pakvasa and
Petrov suggest that D$_{sJ}$ states can be explained by a mixing
of conventional p-wave c$\overline s$ states with 4-quark
states~\cite{Browder}. Since DK or D$\pi$ bound states could have
I=1, one should observe isospin partners of these states. CDF
looks for such isospin partners in D$_s\pi^-$ and D$_s\pi^+$
systems, without finding any peak near 2.32~GeV~\cite{cdf}. A
further hypothesis is proposed by van~Beveren and Rupp~\cite{van
beveren}, who interpret $D_{sJ}^*(2317)$ as a quasi-bound scalar
$c\overline{s}$ state in a unitarized meson model taking into
account the virtual DK scattering channel.

A way to try to identify the observed resonances is to compute
their widths in the hypothesis that they are $c\overline q$ states
and compare the resulting predictions with the experimental
values. In the following, we describe such a calculation. We
should obtain widths of several hundreds MeV for non strange
states and widths less than experimental resolution for states
with strangeness. Furthermore for the latter we should obtain
ratios of radiative and strong widths consistent with experimental
measurements~\cite{Colangelo-Fulvia}.

\section{Strong decays}
Low energy interactions between heavy mesons and light
pseudoscalar mesons can be described by means of HQET and chiral
perturbation theory. In the framework of HQET mesons containing a
single heavy quark are classified in doublets the members of which
are degenerate in mass. Introducing one field for each doublet, it
is possible to build a lagrangian invariant respect to spin and
flavour heavy quark transformations and respect to chiral
transformations for the pseudo Goldstone K, $\pi$ and $\eta$
bosons~\cite{Wise}:
\begin{equation} {\mathcal L} =igTr\{\overline{H}_aH_b\gamma_{\mu}\gamma_5{\mathcal A}_{ba}^{\mu}\}+ihTr\{S_b\gamma_{\mu}\gamma_5{\mathcal A}^{\mu}_{ba}\overline{H}_a\}+i\frac{h'}{\Lambda_{\chi}}Tr\{T^{\mu}_b\gamma_{\lambda}\gamma_5(D_{\mu}{\mathcal A}^{\lambda})_{ba}\overline{H}_a\}+h.c.+... \label{lagr} \end{equation}

where H, S and T represent $\frac{1}{2}^-$, $\frac{1}{2}^+$ and
$\frac{3}{2}^+$ doublets, respectively. These fields are defined
by the following expressions:
\begin{equation}
H_{a}=\frac{1+\not{v}}{2}\left[ P_{a\mu }\gamma ^{\mu }-P_{a}\gamma _{5}%
\right]  \label{H}
\end{equation}

\begin{equation}
S_{a}=\frac{1+\not{v}}{2}\left[ P_{1a}^{\mu }\gamma _{\mu
}\gamma_5-P_{0a} \right]  \label{H}
\end{equation}

\begin{equation}
T^{\mu }=\frac{1}{2}\left( 1+\not{v}\right) \left[ D_{2}^{\mu \nu
}\gamma _{\nu }-\sqrt{\frac{3}{2}}\widetilde{D}_{1\nu }\gamma
_{5}\left( g^{\mu \nu }-\frac{1}{3}\gamma ^{\nu }\left( \gamma
^{\mu }-v^{\mu }\right) \right) \right],  \label{T}
\end{equation}
where $v$ is the meson four-velocity and $a$ is a light quark
flavour index. Light meson fields are included in
Lagrangian~(\ref{lagr}) through ${\mathcal
A}^{\mu}=\frac{1}{2}(\xi^\dagger
\partial^{\mu}\xi-\xi
\partial^{\mu}\xi^\dagger)$, where
$\xi=exp(\frac{i\tilde{\pi}}{f})$ and

\begin{equation}
\widetilde{\pi }=\left(
\begin{array}{ccc}
\frac{\pi _{0}}{\sqrt{2}}+\frac{\eta }{\sqrt{6}} & \pi ^{+} & K^{+} \\
\pi ^{-} & -\frac{\pi _{0}}{\sqrt{2}}+\frac{\eta }{\sqrt{6}} & K^{0} \\
K^{-} & \overline{K}^{0} & -\frac{2}{\sqrt{6}}\eta%
\end{array}
\right) .  \label{pi tilde}
\end{equation}

In our discussion we consider the term of Lagrangian~(\ref{lagr})
describing the coupling of a state of the $\frac{1}{2}^+$ doublet
with one of the $\frac{1}{2}^-$ doublet and a light pseudoscalar
meson. Such interaction is characterized by the coupling constant
$h$, for which we use a value calculated in the framework of QCD
sum rules: $|h|=0.6\pm0.2$~\cite{h}.

Due to parity and angular momentum conservation and consistently
with the experimental masses, the only two-body strong decays
allowed for the non strange 0$^+$ and 1$^+$ states are those in
$D\pi$ and $D^*\pi$, respectively. Considering the average value
of the masses measured by Belle and FOCUS, we obtain: $\Gamma
\left(D_0^{*0}\rightarrow D^+\pi^-\right)$=260$\pm 54$~MeV and
$\Gamma \left(D_1'\rightarrow D^{*+}\pi^-\right)$=160$\pm 25$~MeV,
where the errors reflect the uncertainties of the computational
scheme. Taking into account also decays in $D^0\pi^0$ and
$D^{*0}\pi^0$ we obtain $\Gamma (D_0^{*0})\simeq \frac{3}{2}\Gamma
\left(D_0^{*0}\rightarrow D^+\pi^-\right)=390\pm 80$~MeV and
$\Gamma (D_1')\simeq \frac{3}{2}\Gamma \left(D_1'\rightarrow
D^{*+}\pi^-\right)=240\pm 40$~MeV.

For $c\overline{s}$ states the derivation of the width is less
direct than in the non strange case, due to the isospin violation
occurring in $D_{sJ}\rightarrow D_s^{(*)}\pi^0$ decays. These
decays can be interpreted through a process in two steps: an
isospin-conserving decay with a virtual $\eta$ in the final state
and a $\pi^0-\eta$ mixing due to the difference in mass between u
and d quarks. In fact the mass term in the low energy Lagrangian
describing $\pi$, K and $\eta$ mesons, ${\mathcal
L}=\frac{\tilde{\mu}f^2}{4}Tr\left[\xi m_q\xi +\xi^\dagger
m_q\xi^\dagger \right]$, gives rise to the following matrix
element, vanishing in the limit $m_u=m_d$:
\begin{equation}
    \left\langle \pi^0| \eta \right\rangle= \frac{\tilde{\mu}}{2}\frac{m_d-m_u}{\sqrt 3}.
    \label{mixing}
\end{equation}

The resulting expression for the $D_{s0}^*\rightarrow D_{s}\pi
^{0}$ width is
\begin{equation}
\Gamma \left( D_{s0}^{\ast }\rightarrow D_{s}\pi ^{0}\right)
=\frac{1}{16\pi}\frac{h^{2}}{f^{2}}\frac{M_{D_{s}}}{M_{D_{s0}^{\ast
}}}\left( m_{\pi ^{0}}^{2}+\left\vert \overline{q}\right\vert
^{2}\right) \left\vert
\overline{q}\right\vert \left( \frac{m_{u}-m_{d}}{\frac{m_{u}+m_{d}}{2}-m_{s}%
}\right) ^{2},  \label{Ds0}
\end{equation}
whereas the $D'_{s1}\rightarrow D_{s}^*\pi ^{0}$ width is
expressed by a relation similar to (\ref{Ds0}), the only
difference consisting in a factor $\frac{1}{3}
\left(2+\frac{\left(M^2_{D_{s1}'}+M^2_{D_s^*}-m^2_{\pi^0}
\right)^2}{4M^2_{D_{s1}'} M^2_{D_s^*}}\right)$. Using the factor
$\frac{m_d-m_u}{m_s-\frac{m_d+m_u}{2}}\simeq
\frac{1}{43.7}$~\cite{gasser}, the results $\Gamma
\left(D_{s0}^*\rightarrow D_s\pi^0\right)$=7$\pm 1$~KeV and
$\Gamma \left(D'_{s1}\rightarrow D_s^*\pi^0\right)$=7$\pm 1$~KeV
can be obtained, so that hadronic widths result of the typical
size of radiative widths.

\section{Mixing between $(1^+)_{\frac{1}{2}}$ and
$(1^+)_{\frac{3}{2}}$ states}

In computing strong widths for the axial states of $c\overline{u}$
and $c\overline{s}$ systems we have implicitly assumed that
(1$^+)_{\frac{1}{2}}$ and (1$^+)_{\frac{3}{2}}$ states do not mix.
Due to the finite mass of charm it is instead possible that
physical states are the result of a mixing through an angle
$\theta$ of HQET states. We could have an evidence of such a
mixing if using for the coupling constant
$\frac{h'}{\Lambda_{\chi}}$ appearing in Lagrangian (\ref{lagr})
the value we deduce from the experimental width of the
(2$^+)_{\frac{3}{2}}$ state we compute a width for the
(1$^+)_{\frac{3}{2}}$ state inconsistent with the experimental
measurements.

\begin{table}[htbp]
    \caption{Masses and widths for $s_l^P=\frac{3}{2}^+$ $c\overline{u}$ states
    recently measured by Belle and FOCUS, together the corresponding quantities reported on Particle Data Group.}
    \label{tab:3/2}
    \begin{center}
    \begin{tabular}{ccccc}
      \hline
      \ &  & Belle~\cite{Bellelarghi} &  FOCUS~\cite{focus} & PDG~\cite{pdg} \\
      \hline
      \ $D_1^0 $ & \begin{tabular}{c} Mass (MeV) \\ Width (MeV) \end{tabular}   &
      \begin{tabular}{c} $2421\pm 1.5 \pm 0.8$ \\ $23.7 \pm2.7\pm 0.2\pm 4.0$ \end{tabular}   &
      \begin{tabular}{c}  - \\ - \end{tabular}  &
      \begin{tabular}{c}  $2422.2\pm1.8$ \\ $18.9^{+4.6}_{-3.5}$ \end{tabular}  \\
      \hline
      \ $D_2^{*0}$ & \begin{tabular}{c} Mass (MeV) \\ Width (MeV) \end{tabular}   &
      \begin{tabular}{c} $2462\pm2.1\pm0.5\pm3.3$ \\ $45.6\pm4.4\pm6.5\pm1.6$ \end{tabular}   &
      \begin{tabular}{c}  $2464.5\pm1.1\pm1.9$ \\ $38.7\pm5.3\pm2.9$ \end{tabular}  &
      \begin{tabular}{c}  $2458.9\pm2.0$ \\ $23\pm5$ \end{tabular}  \\
      \hline
    \end{tabular}
  \end{center}
\end{table}

In fitting $D^{**}$ resonances Belle and FOCUS have measured
masses and widths for the narrow $\frac{3}{2}^+$ states different
from those reported on PDG, as it is shown in Table~\ref{tab:3/2}.
The previous experimental measurements could be affected by
systematical errors due to the neglect of the broad resonances.
The experimental value for $\frac{h'}{\Lambda_{\chi}}$ deduced
using these new measurements is $(0.72\pm0.06)\times 10^{-3}$.
Using this experimental value we obtain a width
$\Gamma(D_1^0)=(17\pm6)$~MeV, which is consistent with Belle
measurement. So we conclude that, using recent data, there is no
evidence of a large mixing\footnote{In ref.~\cite{Bellelarghi} a
mixing angle of $\theta\simeq-0.10~rad$ is estimated.}.

\section{Radiative decays}
The amplitude of radiative decays of mesons containing a single
heavy quark can be determined through a method based on the use of
heavy quark symmetries together with the vector meson dominance
(VMD) ansatz.

Computing the coupling of the photon to the heavy quark part of
the e.m. current in the heavy quark limit one obtains that the
matrix element ${\bra
{D^*(v',\epsilon)}}\overline{c}\gamma_{\mu}c{\ket{D_0^*(v)}}$ is
proportional to the inverse heavy quark mass, so that this
contribution can be neglected. Invoking the VMD ansatz the
coupling of the photon with the light quark can be expressed
through an intermediate light vector meson, for example
$\phi$(1020) in the case of heavy mesons with strangeness:
\begin{eqnarray}
\left\langle D_{s}^*\left( p,\epsilon \right) \gamma \left(
q,\widetilde{\epsilon }\right) \mid D_{s0}^*\left( p+q\right)
\right\rangle &\simeq &ee_{s}\sum_{\lambda }\left\langle
D_{s}^*\left( p,\epsilon \right) \phi \left( q,\eta ^{\left(
\lambda \right) }\right) \mid D_{s0}^*\left( p+q\right)
\right\rangle \frac{i}{q^{2}-m_{\phi }^{2}}\cdot
\nonumber \\
&&\cdot \left\langle 0\mid \overline{s}\gamma ^{\mu }s\mid \phi
\left(
q,\eta ^{\left( \lambda \right) }\right) \right\rangle \widetilde{%
\epsilon }_{\mu }^*.   \label{auto}
\end{eqnarray}%

The interaction of a light vector meson with two heavy mesons can
be described by means of a Lagrangian derived in~\cite{simm gauge
nasc} through the hidden gauge symmetry method:
\begin{equation} {\mathcal L'} =i\hat{\mu}Tr\{\overline{S}_aH_b\sigma^{\lambda\nu}V_{\lambda\nu}(\rho)_{ba}\}+h.c. \label{mesonivettori} \end{equation}
where
$V_{\lambda\nu}(\rho)=\partial_{\lambda}\rho_{\nu}-\partial_{\nu}\rho_{\lambda}+[\rho_{\lambda},\rho_{\nu}]$
and $\rho_{\lambda}=i\frac{g_V}{\sqrt{2}}\hat{\rho}_\lambda$,
$\hat{\rho}_\lambda$ being a 3$\times$3 matrix analogue to
$\tilde{\pi}$ defined in~(\ref{pi tilde}) and $g_V$ being fixed to
$g_V=5.8$ in \cite{gV}. Using for $\hat{\mu}$ the value deduced
through the analysis of the $D\rightarrow K^*$ semileptonic
transitions in \cite{mu hat}, we obtain for states with
strangeness: $\Gamma \left(D_{s0}^*\rightarrow
D_s^*\gamma\right)\simeq 1$~KeV, $\Gamma \left(D'_{s1}\rightarrow
D_s\gamma\right)\simeq 3.3$~KeV and $\Gamma
\left(D'_{s1}\rightarrow D_s^*\gamma\right)\simeq 1.5$~KeV.

Our results for ratios between strong and radiative decay rates
for these narrow states are shown in Table~\ref{tab:frazioni}
together the corresponding data of Belle, BaBar and CLEO
Collaborations. The most meaningful comparison is that for the
only observed
radiative decay: our result for $\frac{\Gamma \left( D_{s1}^{\prime }\rightarrow D_{s}\gamma \right) }{%
\Gamma \left( D_{s1}^{\prime }\rightarrow D_{s}^{\ast }\pi
^{0}\right) }$ is consistent with both Belle and BaBar
measurements.

\begin{table}[h]
\caption{Decay fractions for $\frac{1}{2}^+$ states with
strangeness. The values labeled by $\left( \star \right) $ are
those obtained from B decays data reported in~\cite{Belle dal
B},~\cite{BaBardalB}. The other ones result from Belle~\cite{Belle
continuo} and CLEO~\cite{CLEO} continuum analysis.}
\label{tab:frazioni}
    \begin{center}
\begin{tabular}{ccccc}
\hline & Belle & BaBar & CLEO & prediction \\
\hline
$\frac{\Gamma \left( D_{s0}^{\ast }\rightarrow D_{s}^{\ast }\gamma \right) }{%
\Gamma \left( D_{s0}^{\ast }\rightarrow D_{s}\pi ^{0}\right) }$ &
\begin{tabular}{c} $\left(
\star \right) \ 0.29\pm 0.26$ $\left( <0.9\right) $ \\ $<0.18$
\end{tabular}
 & \ \hfill --- \hfill\ & $<0.059$ & $0.1$ \\
\hline
$\frac{\Gamma \left( D_{s1}^{\prime }\rightarrow D_{s}\gamma \right) }{%
\Gamma \left( D_{s1}^{\prime }\rightarrow D_{s}^{\ast }\pi
^{0}\right) }$ &
\begin{tabular}{c} $\left(
\star \right) \ 0.38\pm 0.11\pm 0.04$  \\ $0.55\pm 0.13\pm 0.08$
\end{tabular} & $\left(
\star \right) 0.44\pm0.17$
&$<0.49$ & $0.5$ \\
\hline $\frac{\Gamma \left( D_{s1}^{\prime }\rightarrow
D_{s}^{\ast }\gamma \right) }{\Gamma \left( D_{s1}^{\prime
}\rightarrow D_{s}^{\ast }\pi ^{0}\right) }$ &\begin{tabular}{c}
$\left( \star \right) \ 0.15\pm 0.11$ $\left( <0.4\right) $  \\
$<0.31$
\end{tabular} & \ \hfill --- \hfill\
&$<0.16$ & $%
0.2$ \\
\hline

\ $\frac{\Gamma \left( D_{s1}^{\prime }\rightarrow D_{s}^{\ast
}\gamma \right) }{\Gamma \left( D_{s1}^{\prime }\rightarrow
D_{s}\gamma \right) }$ & $\left(
\star \right) $ $0.40\pm 0.28$ $\left( <1.1\right) $ & \ \hfill --- \hfill\ & \ \hfill --- \hfill\ & $0.4$ \\
\hline
\end{tabular}
\end{center}
\end{table}

Using the same method, we have obtained for the radiative decays
of $c\overline{u}$ and $c\overline{d}$ states the following rates:
$\Gamma \left(D_0^{*0}\rightarrow D^{*0}\gamma\right)= 26\pm
4$~KeV, $\Gamma \left(D_0^{*+}\rightarrow D^{*+}\gamma\right)=
7\pm 1$~KeV, $\Gamma \left(D_1'^+\rightarrow D^+\gamma\right)=
13\pm 3$~KeV, $\Gamma \left(D_1'^0\rightarrow D^0\gamma\right)=
50\pm 10$~KeV, $\Gamma \left(D_1'^+\rightarrow
D^{*+}\gamma\right)\simeq 7$~KeV, $\Gamma \left(D_1'^0\rightarrow
D^{*0}\gamma\right)\simeq 27$~KeV.

\section{Predictions for beauty mesons belonging to $\frac{1}{2}^+$ doublet}
The method applied for c$\overline{q}$ states of the
$\frac{1}{2}^+$ doublet can be also used for the analogue
b$\overline{q}$ states. Since there is no experimental evidence
for the latter, at first we have to estimate their masses.

In the HQET framework the masses of heavy mesons belonging to the
$\frac{1}{2}^-$ doublet are expressed through the following
relation:

\begin{equation}
  m_M=m_Q+\overline{\Lambda}+\frac{\Delta m_M^2}{2m_Q}+... ,
  \label{massemesoni}
\end{equation}

where M=P, V (pseudoscalar, vector). In eq.~(\ref{massemesoni})
the parameter $\overline{\Lambda}$ is independent of heavy quark
flavour and spin, while the correction $\frac{\Delta m_M^2}{2m_Q}$
can be parameterized through the relation $\Delta
m^2_M=-\lambda_1+d_M\lambda_2$, where $\lambda_1$ and $\lambda_2$
are related, respectively, to the matrix elements of the kinetic
and chromomagnetic operators appearing as ${\mathcal
O}(\frac{1}{m_Q})$ corrections to the HQET Lagrangian. Since
$d_P=-3$ and $d_V=1$, one can consider the spin averaged mass
$\tilde{m}=\frac{m_P+3m_V}{4}=m_Q+\overline{\Lambda
}-\frac{\lambda_1}{2m_Q}$, and the analogue expression
$\tilde{m}^*$ for the $\frac{1}{2}^+$ doublet. Disregarding the
$\frac{\lambda_1'-\lambda_1}{2m_Q}$ term in both charm and beauty
cases, we obtain the relation
\begin{equation}
  \tilde{m}^*_{[b]}=\tilde{m}_{[b]}+\tilde{m}^*_{[c]}-\tilde{m}_{[c]},
  \label{centridimassa}
\end{equation}
by means of which we predict for $\frac{1}{2}^+$ beauty mesons
masses the values shown in Table~\ref{tab:beauty}. The $B_{s0}^*$
and $B_{s1}'$ masses are predicted to be below BK and B$^*$K
thresholds, respectively, and so these states are expected to be
detected as narrow resonances in the B$_s\pi^0$ and B$_s^*\pi^0$
systems, analogously to the charm case. Using these mass
predictions we have computed strong and radiative decay rates,
obtaining total widths reported in Table~\ref{tab:beauty} and
decay fractions shown in Table~\ref{tab:beauty
fractions}~\cite{nostro}. Other predictions concerning the masses
of $\frac{1}{2}^+$ $\overline{b}q$ mesons are collected
in~\cite{nostro}, see also~\cite{eef}.

\begin{table}[h]
    \caption{Masses and widths predicted for $\overline{b} q$ states belonging to  $\frac{1}{2}^+$ doublet.}
    \label{tab:beauty}
    \begin{center}
    \begin{tabular}{ccc}
\hline  meson & mass (MeV) & width (MeV) \\
\hline
$B_{0}^{\ast 0}$ & $5710$ & $330\pm 24$ \\
$B_{s0}^{\ast }$ & $5721$ & $\left( 10.5\pm 0.5\right) \times 10^{-3}$ \\
$B_{1}^{\prime 0}$ & $5744$ & $204\pm 14$ \\
$B_{s1}^{\prime }$ & $5762$ & $\left( 11\pm 0.5\right) \times 10^{-3}$ \\
\hline
    \end{tabular}
     \end{center}
\end{table}

\begin{table}[h]
    \caption{Decay fractions predicted for $\overline{b} s$ states belonging to $\frac{1}{2}^+$ doublet.}
    \label{tab:beauty fractions}
    \begin{center}
\begin{tabular}{cc}
\hline  & prediction \\
\hline
$\frac{\Gamma \left( B_{s0}^{\ast }\rightarrow B_{s}^{\ast }\gamma \right) }{%
\Gamma \left( B_{s0}^{\ast }\rightarrow B_{s}\pi ^{0}\right) }$ & $0.4$ \\
\hline
$\frac{\Gamma \left( B_{s1}^{\prime }\rightarrow B_{s}\gamma \right) }{%
\Gamma \left( B_{s1}^{\prime }\rightarrow B_{s}^{\ast }\pi
^{0}\right) }$
& $0.3$ \\
\hline $\frac{\Gamma \left( B_{s1}^{\prime }\rightarrow
B_{s}^{\ast }\gamma \right) }{\Gamma \left( B_{s1}^{\prime
}\rightarrow B_{s}^{\ast }\pi ^{0}\right) }$
 & $%
0.3$ \\
\hline
\end{tabular}
\end{center}
\end{table}

\section{Conclusions}

To test the interpretation of the recently observed resonances as
$c\overline{q}$ states of the $\frac{1}{2}^+$ doublet we have
computed their strong and radiative decay rates in this
hypothesis, using experimental masses. For non strange states we
have obtained widths of several hundreds MeV, consistently with
the measurements, whereas for states with strangeness strong decay
rates are obtained of the same order of the radiative ones, so
that their total widths result to be very narrow, in agreement
with
the experimental observations. For the ratio $\frac{\Gamma \left( D_{s1}^{\prime }\rightarrow D_{s}\gamma \right) }{%
\Gamma \left( D_{s1}^{\prime }\rightarrow D_{s}^{\ast }\pi
^{0}\right) }$ our result is consistent with both measurements
performed by Belle. Ratios of the same order are obtained for the
radiative decays $D_{s0}\rightarrow D_s^* \gamma$ and
$D_{s1}\rightarrow D_s^* \gamma$, though such processes have not
been observed yet. Only upper limits at 90$\%$~C.~L. are
available, with which our results are in substantial agreement, as
shown in Table \ref{tab:frazioni}. So we conclude that the
experimental observations are compatible with the interpretation
of these resonances as $c\overline{q}$ states of the
$\frac{1}{2}^+$ doublet, although the question concerning the low
masses of the states with strangeness remains an open issue. This
interpretation, also proposed in~\cite{godfrey}-\cite{bardeen},
could be corroborated by a future observation of the other
expected radiative decays.

Finally, we have predicted masses and widths for $b\overline{q}$
states of the $\frac{1}{2}^+$ doublet. For $b\overline{s}$ states,
we obtain a scenario similar to the corresponding one in the charm
case, so that we expect that $B_{s0}^*$ and $B_{s1}'$ can be
observed as narrow resonances in the $B_s\pi^0$ and $B_s^*\pi^0$
systems. Too heavy to be produced at B-factories, such resonances
could be discovered in hadronic experiments like CDF, or through a
LEP data reanalysis.

\vspace{1cm} {\bf Acknowledgements}

I thank Pietro Colangelo, Fulvia De Fazio and Giuseppe Nardulli
for collaboration on the topics discussed here, and Donato Creanza
for interesting discussions. This work is supported by MIUR funds.

\end{document}